\pdfoutput=1
\documentclass[12pt]{iopart}

\usepackage{iopams}
\usepackage{siunitx}
\usepackage{graphicx}
\usepackage{subfigure}
\usepackage[hyphens]{url}

\begin{document}

\title[Comment on Philippe D'Arco et al. 2013]{Comment on 'Symmetry and random sampling of symmetry independent configurations for the simulation of disordered solids'. }

\author{K Okhotnikov\footnote{Present address:
R\&D department, RoboCV, Moscow, 115230, Russia}}

\address{NIMBE, CEA Saclay, 91191 Gif-sur-Yvette, France}
\ead{kirill.okhotnikov@gmail.com}

\vspace{10pt}
\begin{indented}
\item[]June 2016
\end{indented}

\begin{abstract}
The proposed algorithm by Philippe D'Arco et al. 2013 J. Phys.: Condens. Matter 25 355401 was not well compared with other possible solutions for symmetry independent configuration (SIC) generation. In this comment, three well known solutions of SIC searching are discussed: exhaustive explorations, pure random sampling and ``symmetry imposed'' approaches. It is shown, that the advantages of the algorithm published by Philippe D'Arco et al. are questionable. 
\end{abstract}

% Uncomment for PACS numbers
%\pacs{00.00, 20.00, 42.10}
%
% Uncomment for keywords
%\vspace{2pc}
%\noindent{\it Keywords}: XXXXXX, YYYYYYYY, ZZZZZZZZZ
%
% Uncomment for Submitted to journal title message
%\submitto{\JPCM}
%
% Uncomment if a separate title page is required
%\maketitle
% 
% For two-column output uncomment the next line and choose [10pt] rather than [12pt] in the \documentclass declaration
% \ioptwocol
%

\section{Introduction}

Publication of the new algorithm assumes that the approach is better than other available solutions, at least for a certain set of cases (input parameters). Therefore, comparison with existing solutions is a crucial part of all methodological papers. The new method will be useful only if it demonstrates better properties for important applications. The common significant properties of different algorithms are accuracy, computational resources requirements, flexibility, generality and complexity of implementation. In this comment, different algorithms for generation of derivative structures will be discussed and compared with symmetry-enhanced sampling (SES) solution, proposed by d'Arco et. al. 

\section{Comparison with existing approaches}

The implementation of the method in CRYSTAL program is closed source and commercial. Therefore, the method comparison is based on original paper and results of the program execution (demo version). The algorithm implementation details are not discussed.

\subsection{Exhaustive explorations method}

Exhaustive explorations of the derivative structures (or ``brute-force'' approach) is a very common method to obtain possible atoms configurations in supercells for fixed and variable compositions. Before 2013 there were at least three programs, which contain various implementation of the method\cite{Grau2007, Hart2008, disCRYSTAL}. The reference method for the comparison will be a lexicographically-ordered processing approach, firstly adopted by Hart and Forcite\cite{Hart2008} to derivative crystal structures generation. A \textit{supercell} program\cite{supercell2016} were used for benchmarking.

From the first point of view, SES method can work with larger supercells, than ``brute-force'' one, but it is not exactly true. SES method allocates an array in random access memory (RAM) which size is proportional to the total number of derivative structures. Therefore the amount of RAM in the computer is an obvious strict limitation for the SES approach. The SES approach needs 4 bytes per each derivative structure. For modern desktop and laptop computers 4 Gb of the memory is an average value. So the system with more than 1 billion of combinations cannot be solved with standard computer memory resources. ``Brut-force'' structures generation algorithm does not have problem with memory, but it requires CPU resources, proportional to number of output structures. One billion of structures can be processed relatively fast\cite{supercell2016}.

There are two structures discussed in the paper: calcite Mn$_x$Ca$_{1-x}$CO$_3$ and spinel MgAl$_x$Fe$_{2-x}$O$_4$. The most complex example is $2\times2\times1$ supercell of calcite with fifty-fifty of Ca and Mn atoms amount. The total number of combinations for the configuration is \num{2704156} (\num{19219} SIC). Exhaustive explorations of all structures can be done in a few seconds\footnote{\textit{Supercell} program\cite{calcite-benchmark}. Dry-run time on Intel\textsuperscript{\textregistered}~Xeon\textsuperscript{\textregistered} X5550 processor.}. The ``stress-test'' of the SES algorithm in much complicated case with \num{601 080 390} total number of combinations fails, but both of implementation of lexicographically-ordered processing method (enumlib, supercell) successfully passed it\cite{supercell2016}.

Although, the exhaustive explorations method complexity is proportional to the number of derivative structures, the number of the structures grows exponentially with the supercell size. The total number of structures can be too high to process on modern computers even for relatively small supercell sizes. For example, calcite compound Mn$_{0.5}$Ca$_{0.5}$CO$_3$ with supercell size $2\times2\times2$ gives the total number of derivative structures $\approx 1.3\cdot 10^{13}$.

\subsection{Random sampling method}
The random sampling (RS) method is not limited by amount of configurations. It can work with arbitrary supercells sizes. Philippe D'Arco et al. insist that their method can sample SIC with equal probability, compare to RS which does not take advantage of symmetry. Meanwhile, the RS algorithm can be simply improved to produce the same result as SES. For each random sampling structure two trivial additional steps are required: (i) calculation a structure symmetry and (ii) accept each structure not absolutely but with probability proportional to the number of symmetry operation of the structure($P\propto N_{symm})$. To protect the result from duplication an extra RAM required. The amount is proportional to the number of sampled structures, but not to the total number of combinations. An example program, which illustrates this algorithm on MgAl$_x$Fe$_{2-x}$O$_4$ spinel compound can be found in \cite{rs-sample}.

Although, the random sampling can work with any system sizes, the ratio between amount of symmetric and non-symmetric structures decreases rapidly with the system size increasing, which lowers the probability of finding the symmetric structures. This problem can be overcome with a method, described in the next section.

\subsection{``Symmetry imposed'' approach}
Strictly speaking, the ``symmetry imposing'' is a preprocessing step before random sampling or exhaustive explorations. The idea of the method if very simple. At first step, the supercell of initial structure with required dimensions is generated. At the next step, some symmetries are imposed to the cell (spacegroup assigned). Assigning a spacegroup to the supercell makes some of the sites symmetry equivalent. Hence following random sampling or ``brute-force'' should process only to non-equivalent group of sites. The approach is illustrated in figure 1.

\begin{figure*}[t]
\caption{Illustration of ``Symmetry imposed'' approach on simple 2D square lattice. (a) Initial cell has one site with four possible equal probability substitutions ("red", "green", "orange", "white"). (b) Supercell $4 \times 4$ has 16 disordered sites. The total number of combinations $C=\frac{16!}{4! \cdot 4! \cdot 4! \cdot 4!}$ is \num{63063000}. (c) The supercell with two symmetry operations imposed (vertical and horizontal mirroring). Total number of symmetry independent sites is 4 and the total number of combinations $C=4!$ is 24.}
\centering
\subfigure[]{
\includegraphics[trim={5cm 1cm 5cm 1cm}, clip, width=0.3\linewidth]{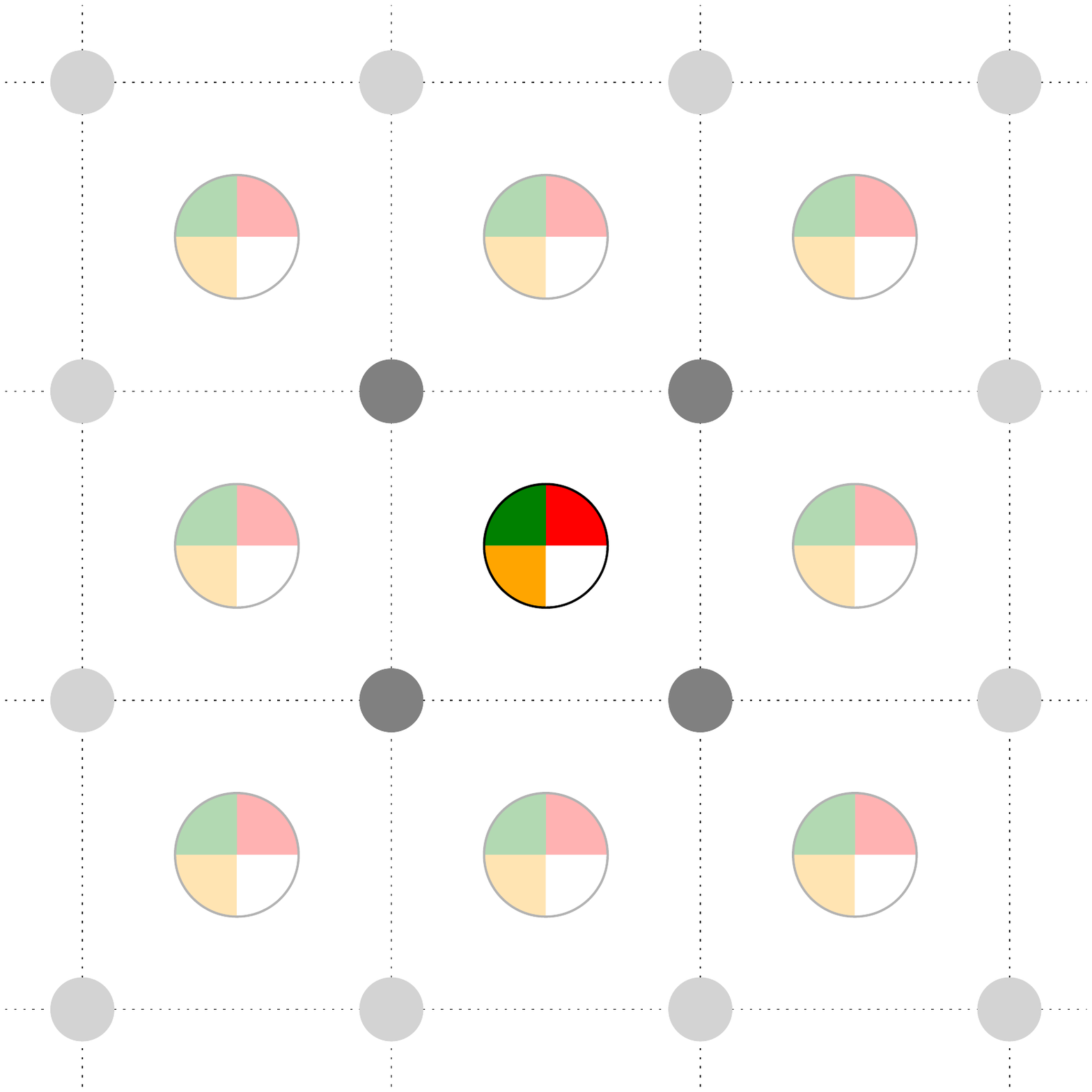}
\label{fig:si:init}
}
\subfigure[]{
\includegraphics[trim={5cm 1cm 5cm 1cm}, clip, width=0.3\linewidth]{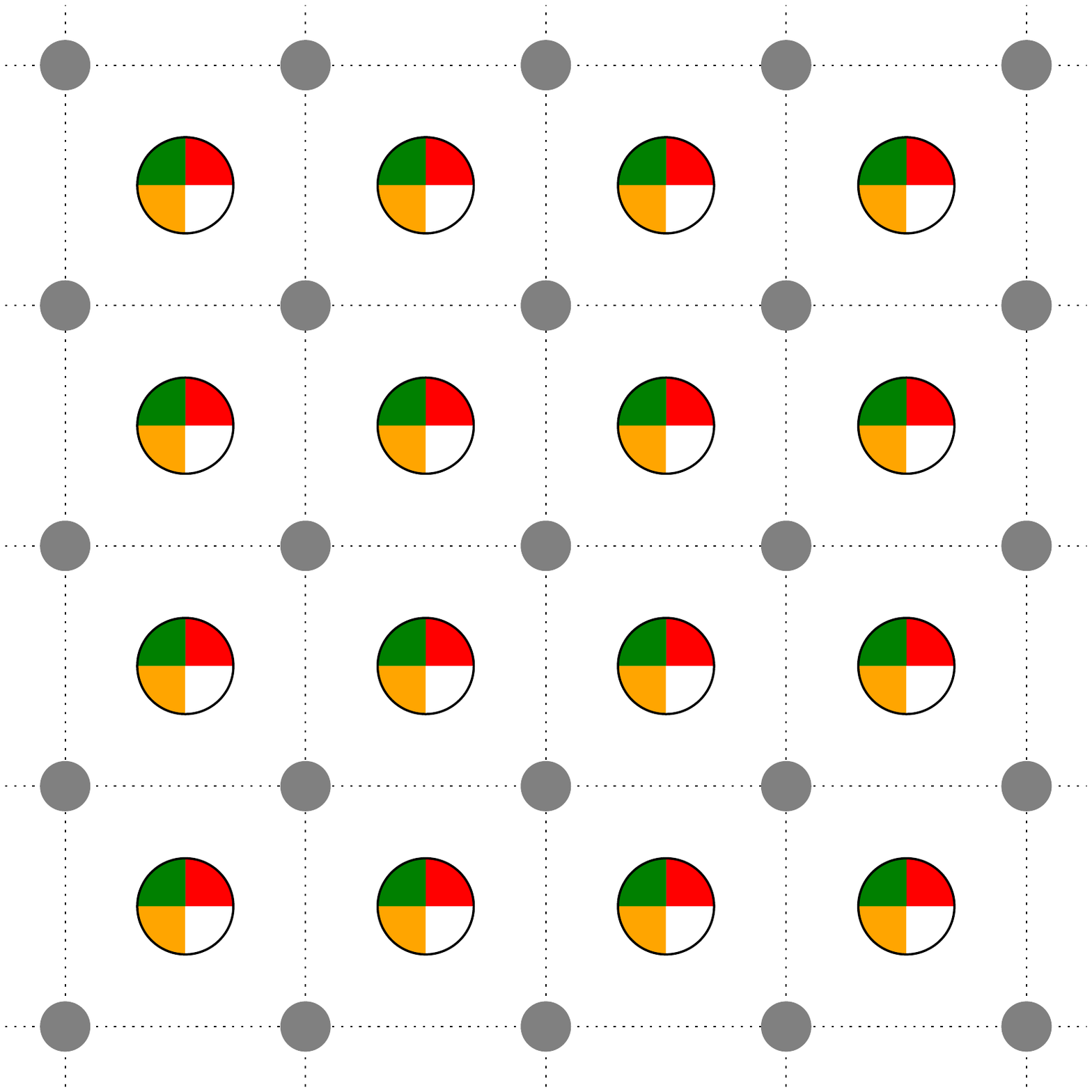}
}
\subfigure[]{
\includegraphics[trim={5cm 1cm 5cm 1cm}, clip, width=0.3\linewidth]{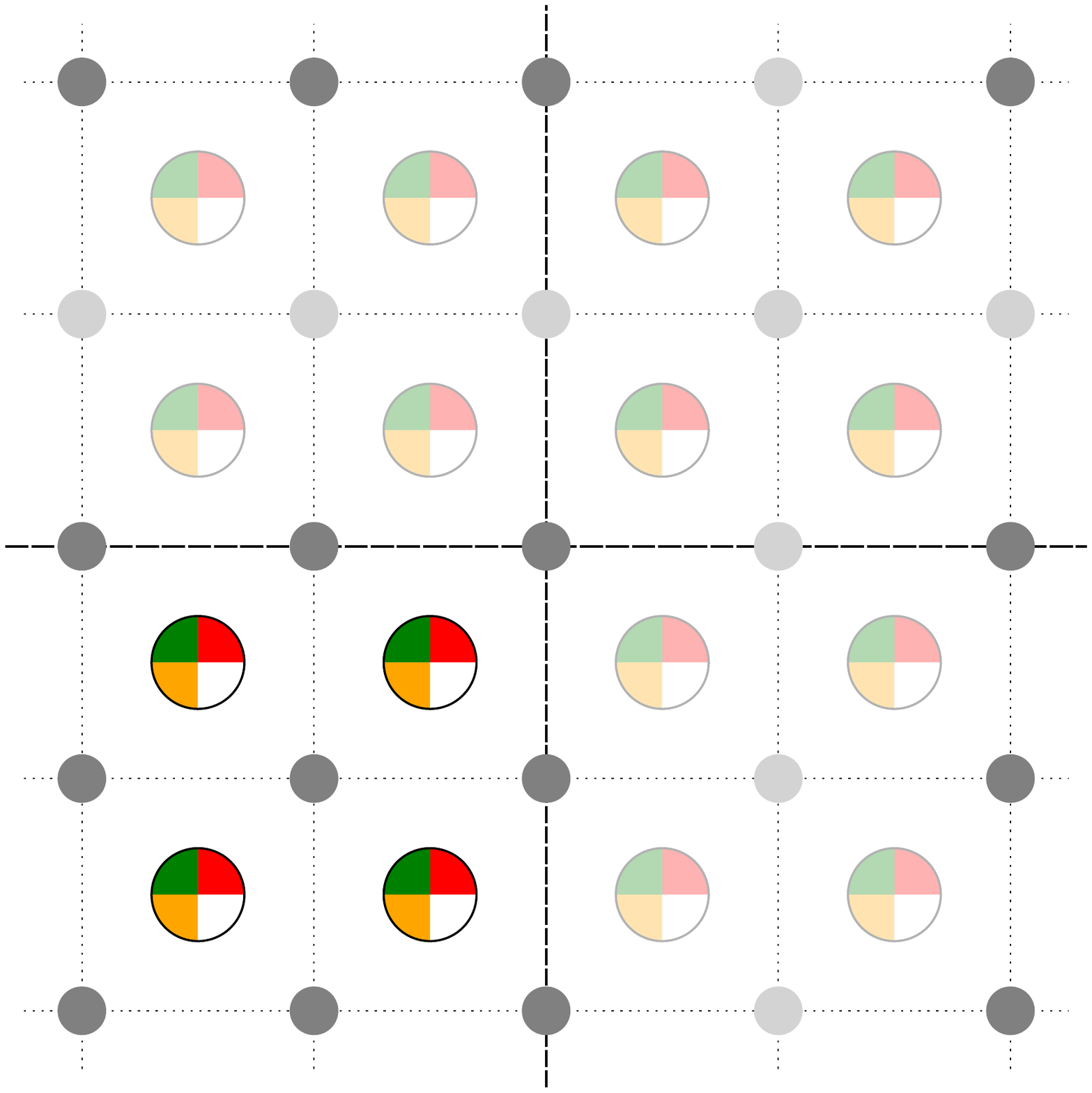}
}
\label{fig:si} 
\end{figure*}   

An impressive example of the approach, can be found in\cite{Paglia2005}. In the paper, the application of symmetry restrictions (in combination with connectivity restrictions) decreases the number of derivative structures from $4.25 \times 10^{12}$ to 513 configurations.

\section{Conclusion}
Although the SES algorithm seems to give a correct solution of disordered structures enumeration, other approaches, existing before, are more flexible (``brute-force''), better scalable (RS, ``Symmetry imposed''), easier to implement and have less limitations.

\section{Appendix}

\section*{References}

\bibliographystyle{iopart-num} 
\bibliography{cmrandom}

\end{document}